\documentclass[journal]{IEEEtran}
\usepackage{graphicx,here}

\begin{document}
\title{\tiny\centering \vspace{-2mm}This work has been submitted to the IEEE Nuclear Science Symposium 2012 for publication in the conference record. Copyright may be transferred without notice, after which this version may no longer be available.\\
\Huge
Beam Test Results for New Full-scale GEM Prototypes for a Future Upgrade of the \\ CMS High-$\eta$ Muon System }

\author{
D.~Abbaneo, M.~Abbrescia, C.~Armagnaud, P.~Aspell, Y.~Assran, Y.~Ban, S.~Bally, L.~Benussi, U.~Berzano, S.~Bianco, J.~Bos, K.~Bunkowski,  J.~Cai, J.~P.~Chatelain, J.~Christiansen, S.~Colafranceschi, A.~Colaleo, A.~Conde~Garcia, E.~David, G.~de~Robertis, R.~De~Oliveira, S.~Duarte Pinto, S.~Ferry, F.~Formenti, L.~Franconi, T.~Fruboes, A.~Gutierrez, M.~Hohlmann,~\IEEEmembership{Member,~IEEE}, A.~E.~Kamel, P.~E.~Karchin, F.~Loddo, G.~Magazz\`u, M.~Maggi, A.~Marchioro, A.~Marinov, K.~Mehta, J.~Merlin, A.~Mohapatra, T.~Moulik, M.~V.~Nemallapudi, S.~Nuzzo, E.~Oliveri, D.~Piccolo, H.~Postema, A.~Radi, G.~Raffone, A.~Rodrigues, L.~Ropelewski, G.~Saviano, A.~Sharma,~\IEEEmembership{Senior Member,~IEEE}, M.~J.~Staib, H.~Teng, M.~Tytgat,~\IEEEmembership{Member,~IEEE}, S.~A.~Tupputi, N.~Turini, N.~Smilkjovic, M.~Villa, N.~Zaganidis, M.~Zientek

\thanks{Manuscript received November 16, 2012.}

\thanks{M.~Abbrescia, A.~Colaleo, G.~de~Robertis, F.~Loddo, M.~Maggi, S.~Nuzzo and S.~A.~Tupputi are with Politecnico di Bari, Universit\`a di Bari and INFN Sezione di Bari - Bari, Italy.}
\thanks{Y.~Ban, J.~Cai and H.~Teng are with Peking University - Beijing, China.}
\thanks{T.~Moulik is with NISER - Bhubaneswar, India.}
\thanks{Y.~Assran, E.~A.~Kamel and A.~Radi are with ASRT-ENHEP - Cairo, Egypt.}
\thanks{A.~Gutierrez and P.~E.~Karchin are with Wayne State Univ. - Detroit, USA.}
\thanks{L.~Benussi, S.~Bianco, D.~Piccolo, G.~Raffone and G.~Saviano are with Laboratori Nazionali di Frascati INFN - Frascati, Italy.}
\thanks{D.~Abbaneo, C.~Armagnaud, P.~Aspell, Y.~Assran, S.~Bally, U.~Berzano, J.~Bos, K.~Bunkowski, J.~P.~Chatelain, J.~Christiansen, S.~Colafranceschi$^*$, A.~Conde~Garcia, E.~David, R.~De~Oliveira, S.~Duarte Pinto, S.~Ferry, F.~Formenti, L.~Franconi, A.~Marchioro, K.~Mehta, J.~Merlin, M.~V.~Nemallapudi, H.~Postema, A.~Rodrigues, L.~Ropelewski, A.~Sharma, N.~Smilkjovic, M.~Villa and M.~Zientek are with Physics~Department,~CERN - Geneva,~Switzerland.}
\thanks{A.~Marinov, M.~Tytgat and N.~Zaganidis are with Department of Physics and Astronomy Universiteit Gent - Gent, Belgium.}
\thanks{M.~Hohlmann$^*$, A.~Mohapatra and M.~J.~Staib are with Florida Institute of Technology - Melbourne, USA.}
\thanks{G.~Magazz\`u, E.~Oliveri and N.~Turini are with INFN Sezione di Pisa - Pisa, Italy.}
\thanks{T.~Fruboes is with Warsaw University - Warsaw, Poland.}
\thanks{$^*$Corresponding authors: stefano.colafranceschi@cern.ch and \newline hohlmann@fit.edu.}
}

\maketitle
\pagestyle{plain}

\begin{abstract}
The CMS GEM collaboration is considering Gas Electron Multipliers (GEMs) for upgrading the CMS forward muon system in the $1.5<|\eta|<2.4$ endcap region. GEM detectors can provide precision tracking and fast trigger information. They would improve the CMS muon trigger and muon momentum resolution and provide missing redundancy in the high-$\eta$ region. Employing a new faster construction and assembly technique, we built four full-scale Triple-GEM muon detectors for the inner ring of the first muon endcap station. We plan to install these or further improved versions in CMS during the first long LHC shutdown in 2013/14 for continued testing. These detectors are designed for the stringent rate and resolution requirements in the increasingly hostile environments expected at CMS after the second long LHC shutdown in 2018/19. The new prototypes were studied in muon/pion beams at the CERN SPS. We discuss our experience with constructing the new full-scale production prototypes and present preliminary performance results from the beam test. We also tested smaller Triple-GEM prototypes with zigzag readout strips with 2 mm pitch in these beams and measured a spatial resolution of 73 $\boldmath{\mu m}$. This readout offers a potential reduction of channel count and consequently electronics cost for this system while maintaining high spatial resolution.
\end{abstract}


\section{Introduction}
\IEEEPARstart{S}{}ince the muon endcap region of the CMS experiment\cite{:2008zzk} at $1.6<|\eta|<2.4$ is currently only instrumented with cathode strip chambers, the CMS GEM collaboration has been studying Gas Electron Multiplier\cite{Sauli:1997qp} (GEM) detectors for a potential upgrade\cite{tytgat_ieee2011,TP-2012-CERN note} in that region. GEM detectors are able to cope with the increasingly hostile future environment in this region and could occupy the place of the originally planned but later descoped Resistive Plate Chambers\cite{Santonico:1994dk}. 

Since 2010, the collaboration has performed several feasibility studies on small\cite{tytgat_ieee} and full-size\cite{colafranceschi_ieee} detectors, showing that GEMs are a technology that can sustain the high-$\eta$ environment.
The collaboration demonstrated the maturity of this technology in two previous iterations\cite{colafranceschi_icatp11} of full-scale detectors that showed good spatial and time resolution, and excellent high-rate capability and radiation hardness. In 2012, the collaboration designed and built four new full-scale GEM detector prototypes using a design close to that anticipated for mass production of the chambers. These were succesfully tested in the laboratory with x-rays and in two test beam campaigns at the H4 beam line at the CERN SPS as described below.

\section{Full-scale detector description}

The full-scale CMS GEM chamber is a trapezoid with dimensions $990~\rm{mm} \times (220-455)~\rm{mm}$ as imposed by the geometry of the vacant high-$\eta$ area in the CMS muon endcap. Each chamber hosts a Triple-GEM detector with a 3/1/2/1~mm (drift, transfer 1, transfer 2, induction) electrode gap configuration. The drift electrode is made from printed circuit board and forms the base plate of the assembly onto which the other parts of the detector are mounted.
The GEM foil production relies on the single-sided mask technology\cite{Villa:2010wj} developed at CERN to overcome the problems with alignment of double masks for large surfaces. The GEM foils ($50~{\rm {\mu m}}$ thick kapton sheet clad with $5~{\rm {\mu m}}$ copper on both sides) are sectorized into 35 high voltage sectors transverse to the strip direction so that each sector has a surface area of about $\rm{100 ~cm^2}$ to limit the discharge probability and energy.
The detector readout board is divided into eight $\eta$-partitions containing 384 readout strips oriented radially along the long side of the detector with a pitch varying from 0.6~mm at the short end to 1.2~mm at the wide end; each $\eta$-partition is subdivided along the $\phi$-coordinate into three readout sectors each with 128 strips.

\begin{figure}[H]
  \includegraphics[width=3.7in]{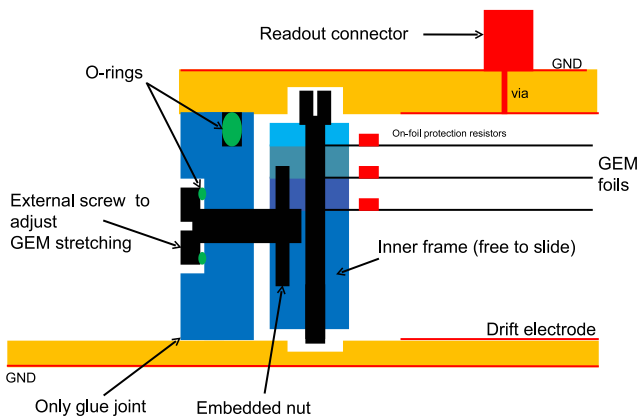}\vspace*{-0.8cm}
  \caption{Cross section of the frame region for a full-size GEM assembled with the new self-stretching technique (not to scale).}
  \label{ns4}
\end{figure}

These new full-scale detectors were assembled using a new technique (Fig.~\ref{ns4}), introduced in 2011, that aims at mechanically stretching GEM foils as part of the assembly process without the use of spacer frames or glue. 
While this new assembly technique was succesfully tested on $10\times10~\rm{cm}^2$ and $30\times30~\rm{cm}^2$ detectors in 2011, it was applied for the first time to full-scale chambers in 2012. The full-scale GEM foils are produced with a pattern of holes along the edges, outside the active area, so that  the foils can be aligned via special pins and fixed to an internal frame. Using screws penetrating the outer frame and fastened by nuts embedded in the inner frame, the foils are stretched against the outer frame when the screws are tightened. The outer frame holds the mechanical tension from the foils and provides gas tightness for the detector. In the final assembly step, the readout board is mounted with screws on top of the outer frame and sealed via an O-ring embedded in a grove in the outer frame, closing the detector. Our experience shows that this assembly procedure can be completed in less than two hours.

\vspace*{-0.3cm}

\section{Test beam setup}
The full-scale CMS detectors were tested during July 2012 with a 150~GeV muon/pion beam at the CERN SPS H4 beam line to validate the new assembly technique applied to these larger chambers.
The test beam setup consisted of a GEM tracking telescope with $\rm{10 \times 10~cm^2}$ active area provided by the RD51 collaboration\cite{RD51} and equipped with 256 strips with 0.4~mm pitch in both horizontal (y-coordinate) and vertical (x-coordinate) directions transverse to the beam. The full-scale CMS detectors were placed on a vertically movable table close to this GEM tracking telescope for scanning.

Both the GEM tracking telescope and the CMS full-scale detectors were read out using VFAT2\cite{Aspell:2008zz} electronics, a digital chip with 128 channels designed with $\rm{0.25~\mu m}$ CMOS technology at CERN using radiation tolerant components. 
The chip, 40~MHz synchronous, features a programmable fast OR function of the input channels for triggering, adjustable threshold and latency with a programmable integration time of the analog input signal. 
During the 2012 test beam campaign, the integration time was set to the minimum value of one clock cycle (25~ns).

\section{Results}
During the test beam campaign the full-scale CMS detectors were operated with $\rm{Ar/CO_2/CF_4}$ 45:15:40. They showed good efficiency, space resolution, and timing performance with stable and reliable operation at gas gains up to $\rm{10^4}$ and with a very low noise level that allowed us to run the electronics with a threshold of $\approx$~0.8~fC.

\begin{figure}
  \centering
  \includegraphics[width=3.5in]{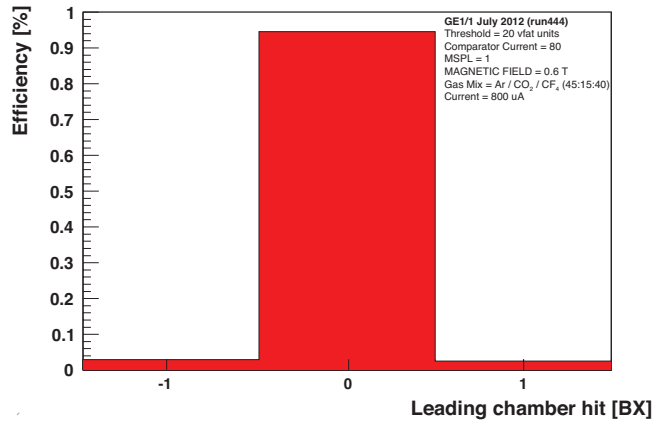}
  \caption{Distribution of 25~ns bunch crossings containing hits in the full-scale CMS detector at B=0.6T using a simulated synchronous beam with 25~ns bunches.}
  \label{ge11timing}
\end{figure}

Fig.~\ref{ge11timing} shows one of the most important timing measurement performed. With the comparator current of the VFAT set to $\rm{80~\mu A}$ and using a newly implemented trigger logic we measured the hit distribution in the 25~ns bunches. The trigger logic was implemented to be synchronous with the chip clock and to reject the muon beam outside a window of $\approx$~25~ns, i.e. simulating the 25~ns bunch structure of the future LHC proton bunch spacing. 
The shown result indicates that $\approx$~95~\% of measured hits in the GEM chamber are contained within one 25~ns bunch crossing at B~=~ 0.6~T.

\begin{figure}
  \centering
  \hspace*{-1mm}\includegraphics[width=3.3in]{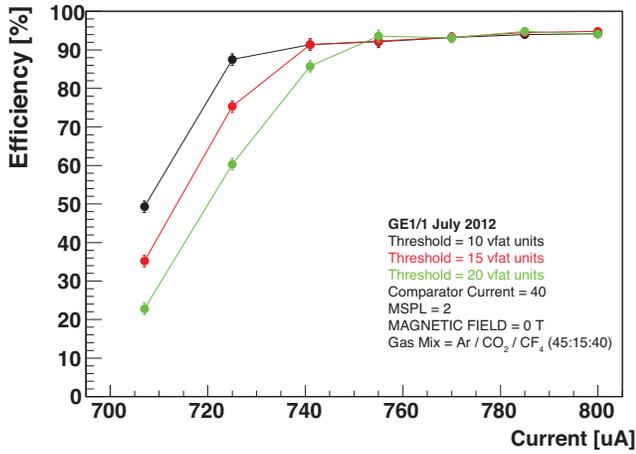}
  \caption{Efficiency vs. HV divider current during high voltage and threshold scan.}
  \label{ge11efficiency}
\end{figure}

\begin{figure}
  \centering
  \includegraphics[width=3.5in]{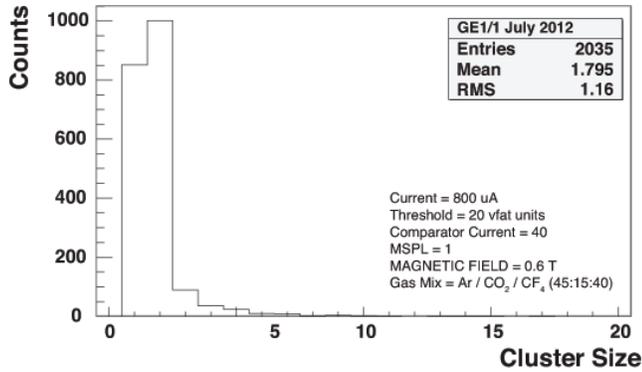}
  \caption{Strip cluster size distribution at B=0.6T when operating chamber on the efficiency plateau.}
  \label{ge11clustersize}
\end{figure}

\begin{figure}[H]
  \centering
  \includegraphics[width=3.4in]{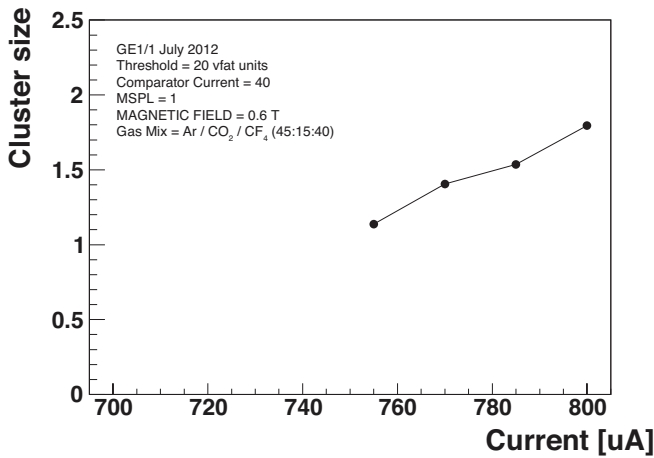}
  \caption{Average strip cluster size vs. HV divider current during HV scan at B=0.6T.}
  \label{ge11clustersizehv}
\end{figure}

Fig.~\ref{ge11efficiency} depicts preliminary results from a high voltage (HV) and threshold scan showing that the detector reaches 95\% efficiency when operated at a HV that corresponds to a gain of $\approx$~7000 in agreement with previous beam test results.
Fig.~\ref{ge11clustersize} and Fig.~\ref{ge11clustersizehv} show the cluster size distribution on the efficiency plateau and the average cluster sizes, respectively, as a function of HV when the chambers are operated in a 0.6~T magnetic field. 

\begin{figure}
  \centering
  \includegraphics[width=3.5in]{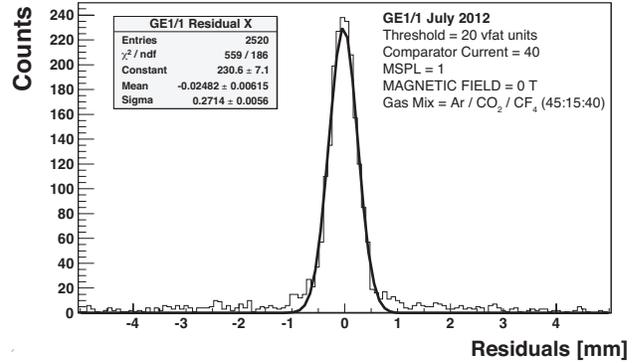}
  \caption{Residuals between the reconstructed tracks and the hits in the full-scale CMS detector at B=0T.} 
  \label{spaceres}
\end{figure}

Fig.~\ref{spaceres} shows the residuals between the beam tracks reconstructed by the GEM telescope and then extrapolated to the position of the full-scale CMS detector and the hit positions measured with the full-scale CMS detector for an $\eta$-section where the strip pitch is $\approx$~0.9~mm. The measured resolution of 270 $\mu m$ is consistent with the $900/\sqrt{12}\ \mu m = 260\ \mu m$ expected for a binary readout.

\section{GEM readout with zigzag strips}
In an effort to significantly reduce overall cost for a full CMS muon endcap system with large-area GEM detectors, we are investigating a readout structure that employs zigzag strips\cite{Wilcox2010} with larger pitch than standard straight strips. Due to the increased pitch, considerably fewer zigzag strips are needed to cover a given readout area compared with straight strips, which reduces the number of required electronic readout channels and ultimately system cost. Due to the interleaving of the strongly slanted zigzag strips, the sharing of induced charge among adjacent strips becomes more sensitive to the position of the electron avalanche along the $\phi$-coordinate, i.e.\ transverse to the main strip direction, than for straight strips. In effect, the increased spatial resolution in that $\phi$-coordinate is being bought by making use of the other (radial) coordinate that is needed to create the 2D zigzag structure. Consequently, zigzag strips in this form are mainly applicable to 1D readouts such as the one planned for the CMS muon endcap system. 

\begin{figure}[H]
  \centering
  \hspace*{2mm}\includegraphics[width=5.6in]{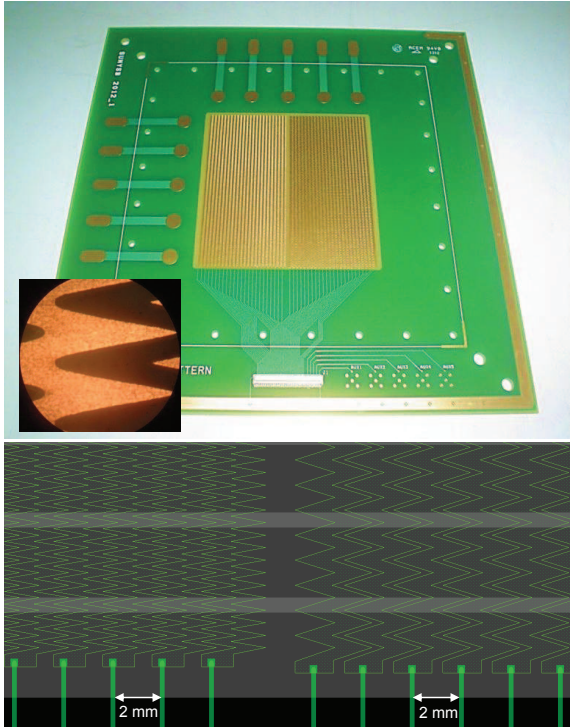}
  \caption{Top: Printed circuit board with 48 zigzag strips with 2~mm strip pitch for a standard CERN 10~$\times$~10~cm$^2$ GEM detector. The inset shows a gold-plated zigzag strip under a microscope. Bottom: Geometry layout of two sets of 24 zigzag strips with different zigzag periods along the strips, i.e.\ 0.5~mm (left) and 1~mm (right).} 
  \label{zigzag-pcb}
\end{figure}

We have developed a small printed circuit readout board featuring 48 zigzag strips with 2~mm strip pitch and 10~cm length (Fig.~\ref{zigzag-pcb}, top) to be used with standard CERN $10\times10~\rm{cm}^2$ Triple-GEM detectors. The strips are divided into two sections of 24 strips each. One section of ``fine'' strips has a 0.5~mm period in the zigzag structure while the ``coarse'' strips have a 1~mm period (Fig.~\ref{zigzag-pcb}, bottom). 

Two such detectors were equipped with this type of pcb and installed as a stacked module in the GEM tracker described above and operated with an Ar/CO$_2$ 70:30 gas mixture in the test beam. The two detectors were placed at the smallest distance from each other allowed by the mounting infra-structure, i.e.\ with 23~mm separating the readout pcbs along the beam direction. As a position measurement using charge-sharing among strips requires pulse height measurements, these detectors were read out independently with the Scalable Readout System\cite{Gnanvo:2010ra,Sorin} developed by the RD51 collaboration with APV25 hybrids at the frontend.

\begin{figure}[H]
  \centering
  \includegraphics[width=3.4in]{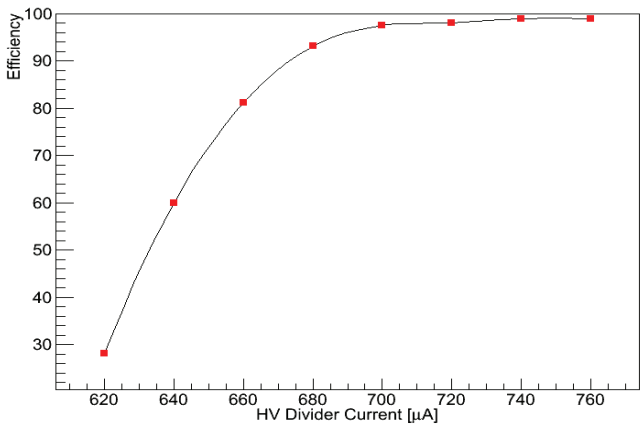}
  \caption{Efficiency vs. HV divider current for 10~$\times$~10~cm$^2$ Triple-GEM detector with zigzag strip readout at B=0T.}
  \label{zigzag-eff}
\end{figure}

\begin{figure}[H]
  \centering
  \includegraphics[width=3.4in]{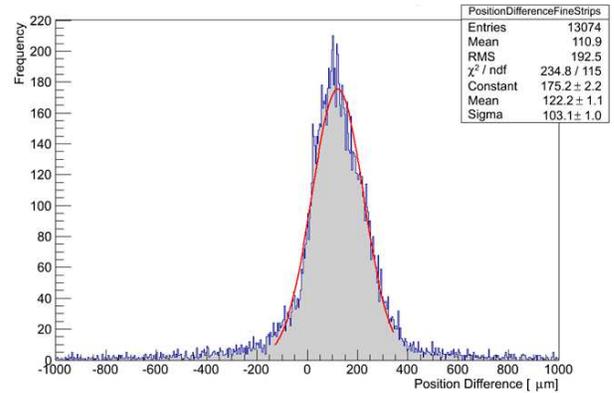}
  \caption{Difference in hit positions measured with two 10~$\times$~10~cm$^2$ Triple-GEM detectors with fine zigzag strip readout placed 23~mm apart (B=0T).} 
  \label{zigzag-res}
\end{figure}

Fig.~\ref{zigzag-eff} shows that the detectors reach 98\% efficiency on plateau. The hit positions for beam particles traversing the two detectors are calculated from the positions of the zigzag strip means in a strip cluster weighted by the measured respective pulse heights (barycenter method). The difference in hit positions measured in the two detectors using fine zigzag strips are histogrammed in Fig.~\ref{zigzag-res}. Dividing the measured rms width of 103~$\mu m$ for that distribution by $\sqrt{2}$ we estimate the resolution of Triple-GEM detectors with zigzag readout to be $\sigma_{zigzag} = 73\ \mu m$.

\section{Summary and Conclusion}
Four full-scale CMS GEM detector prototypes were designed, built, and successfully tested during 2012.
The CMS GEM collaboration implemented a faster assembly technique in the construction of these full-scale dectors. The GEM foils of the new full-scale CMS detectors can be stretched without the need for any spacers and the detectors are constructed only with minimal gluing. The detectors perform well with respect to detection efficiency and spatial resolution as measured in dedicated beam tests during July 2012. A new direct timing measurement was performed showing $\approx$~95\% efficiency in a single 25~ns bunch crossing. We conclude that the faster assembly technique is a viable option that can facilitate mass production of these chambers for the CMS upgrade. A new readout structure for GEM detectors featuring zigzag strips was also successfully tested in the beam. Despite a 2~mm strip pitch, a spatial resolution of 73 $\mu m$ was achieved in small Triple-GEM detectors. We conclude that this readout approach merits further investigation as it could potentially allow a reduction in the number of needed readout strips by roughly a factor of three and consequently save significant electronics cost for the full CMS muon endcap system.

\section{Acknowledgment}
This work has partly been performed in the framework of the RD51 collaboration. We thank the RD51 collaboration for its technical support. We thank T. Hemmick and C. Pancake from SUNY Stony Brook for their help with designing and constructing the zigzag pcb.



\begin{thebibliography}{1}

\bibitem{:2008zzk}
CMS Collaboration,
  ``The CMS experiment at the CERN LHC,''
  JINST {\bf 3} (2008) S08004.

\bibitem{Sauli:1997qp}
F.~Sauli,
  ``GEM: A new concept for electron amplification in gas detectors,''
  Nucl.\ Instrum.\ Meth.\  A {\bf 386}, 531 (1997).

\bibitem{tytgat_ieee2011}
M.~Tytgat {\it et al.},
  ``Construction and Performance of Large-Area Triple-GEM Prototypes for Future Upgrades of the CMS Forward Muon System,'' in Proc.\ IEEE NSS-MIC 2011, N19-7, pp.\ 1019-1025.

\bibitem{TP-2012-CERN note}
D.~Abbaneo {\it et al.} (GEM Coll.\ - GEMs for CMS),
  ``Technical Proposal - A GEM Detector System for an Upgrade of the CMS Muon Endcaps,''
  CMS note, CMS IN 2012/11, CERN.
  
\bibitem{Santonico:1994dk}
R.~Santonico,
  ``RPC: Status and perspectives,''
in Proc., The resistive plate chambers in particle physics and astrophysics, Pavia 1993, 1-11.

\bibitem{tytgat_ieee}
D.~Abbaneo {\it et al.},
  ``Characterization of GEM Detectors for Application in the CMS Muon Detection System,'' in Proc.\ IEEE NSS-MIC 2010, N48-210, pp.\ 1416-1422, 10.1109/NSSMIC.2010.5874006.

\bibitem{colafranceschi_ieee}
D.~Abbaneo {\it et al.},
  ``Construction of the first full-size GEM-based prototype for the CMS high-$\eta$ muon system,'' in Proc.\ IEEE NSS-MIC 2010, N69-1, pp.\ 1909-1913, 10.1109/NSSMIC.2010.5874107.

\bibitem{colafranceschi_icatp11}
D.~Abbaneo {\it et al.},
  ``Construction and Performance of full scale GEM prototypes for future upgrades of the CMS forward Muon system,'' in Proc.\ ICATP 2011.

\bibitem{Villa:2010wj} 
  M.~Villa, {\it et al.},
  ``Progress on large area GEMs,'' Nucl.\ Instrum.\ Meth.\ A {\bf 628}, 182 (2011), arXiv:1007.1131 [physics.ins-det].

\bibitem{RD51}
  The RD51 Collaboration, http://rd51-public.web.cern.ch/RD51-Public/
  
\bibitem{Aspell:2008zz}
P.~Aspell, {\it et al.},
  ``The VFAT production test platform for the TOTEM experiment,'' in Proc.\ Topical Workshop on Electronics for Particle Physics 2008, CERN-2008-008.

\bibitem{Wilcox2010}
R.~Wilcox, B.~Azmoun, C. Woody, ``A Study of Readout Pads for a GEM Detector,'' Rep.\ BNL SULI program 2010.

\bibitem{Gnanvo:2010ra} 
K.~Gnanvo, {\it et al.},
  ``Detection and Imaging of High-Z Materials with a Muon Tomography Station Using GEM Detectors,'' in Proc.\ IEEE NSS-MIC 2010, N19-90, pp.\ 552-559, arXiv:1011.3231 [physics.ins-det]. 

\bibitem{Sorin}
S.~Martoiu {\it et al.},
  ``Front-end electronics for the Scalable Readout System of RD51,'' in Proc.\ IEEE NSS-MIC 2011, N43-5, pp.\ 2036-2038.
  
  

\end{thebibliography}
\end{document}